\documentclass{aa}                             

\newcommand{\be}{\begin{equation}}
\newcommand{\ee}{\end{equation}}
\newcommand{\bea}{\begin{eqnarray}}
\newcommand{\eea}{\end{eqnarray}}
\newcommand{\p}{\partial}

\begin{document} 


\title{Near-integrability as a numerical tool in solar system dynamics}

\author{   M. Kaasalainen    
\and       T. Laakso        } 
\offprints{Mikko.Kaasalainen@astro.helsinki.fi }   
\institute{Observatory, P.O. Box 14, FIN-00014 University of Helsinki,
           Finland\\
           Tel: +358-9-19122802, Fax: +358-9-19122952, Email:
           Mikko.Kaasalainen@astro.helsinki.fi}
\date{Received; accepted} 
\titlerunning{Near-integrability as a numerical tool    } 
\authorrunning{Kaasalainen and Laakso}                    
\abstract{
We present a simple choice of integration variables that can be used
to exploit the near-integrable character of problems in celestial
mechanics. The approach is based on the well-known principle of
variation of parameters: instead of orbital elements, we use
the phase-space coordinates the object would have at a given point
in its (Keplerian) orbit if the perturbing forces were removed.
This formulation is suitable 
for almost any numerical integrator; thus, multistep schemes are easy
to build, stepsize can be adjusted, and dissipative forces are
allowed. Compared with traditional non-symplectic N-body integrators, 
the approach often offers increase in speed or accuracy if perturbations
are small. 
}                              
\maketitle     

\section{Introduction}

Objects in sparsely populated systems dominated by a single massive body 
spend most of their time in the perturbed two-body
state.
Since Euler and Lagrange, dynamicists have constructed
integration methods that can explicitly take into account 
this near-integrable character of problems in celestial mechanics. 
In principle, any scheme that returns the exact Kepler orbit of 
a two-body problem when perturbations are removed describes the
system much better than a `blind' conventional N-body method
(such as one of Gauss-Jackson or other multistep and 
double-integration type),
the foremost advantage being a longer timestep.

No single integration method is automatically superior
to others, owing to the fact that different problems usually
require somewhat different approaches. However,
traditional schemes modelled in cumbersome forms and variables
have lately been replaced by symplectic integrators (SIs):
in addition to allowing the representation of near-integrability, they
exhibit no secular growth of energy error. Development
in this field has been rapid in recent years, and some of the
disadvantages of early SIs have been alleviated by, 
e.g., symplectic correctors, (limited) adjustability of stepsize, and
the possibility to accommodate weak dissipative forces (see, e.g.,
Levison \& Duncan \cite{Le94}; 
Saha \& Tremaine \cite{Sa94}; Wisdom, Holman \& Touma \cite{Wi96}; 
Mikkola \cite{Mi97}, \cite{Mi98}). Symplectic integrators that allow
close encounters have also been constructed
(Duncan, Levison \& Lee \cite{Du98}; Chambers \cite{Ch99}; 
Mikkola \& Tanikawa \cite{Mi99}; Preto \& Tremaine \cite{Pr99};
Levison \& Duncan \cite{Le00}).
However, SIs cannot by definition
tackle general non-Hamiltonian forces, and there is as yet no proper way of
using multistep information to build inexpensive high-order schemes.

The principal manifestation of traditional methods is analytical
perturbation theory (often referred to as `general perturbations'). 
In the numerical domain (`special perturbations') 
most of the traditional schemes have now little more than historical
interest: they were developed for pen and paper, not for modern computing
machines. However, the method known as `variation of parameters'
or `variation of arbitrary constants', used in different forms by
Euler, Lagrange, Poisson and many others after them 
(see, e.g., Herrick \cite{He72}; Danby \cite{Da87}),
is quite useful in its basic principle. The main question is the choice
of parameters, which we discuss in this paper. The geometric
Keplerian elements (and their variants) have usually been the
first choice for variational formulation; however, they are not the
best option for modern purposes. 

We describe an approach that uses
the phase-space coordinates the object would have at a given point
in its (Keplerian) orbit if the perturbing forces were removed.
This results in a simple near-integrable 
formulation that is suitable 
for almost any numerical integrator; one is thus free to
build multistep or hybrid schemes, vary the stepsize, 
and add dissipative forces. This scheme is, in a way, 
complementary to SIs, offering an increase
in speed or accuracy in problems of celestial mechanics where SIs cannot
be employed. 

The basic principles and concepts are presented in Sect.\ 2. In Sect.\ 3
we describe a choice of frame in which  
low-order methods are especially simple to integrate.
In Sect.\ 4 we define
another frame; in this case, any high-order multistep scheme can be 
efficiently applied.  
In Sect.\ 5 we discuss numerical results, and Sect.\ 6 sums up.

\section{Perturbative formulation}

Our goal is to seek elements $\bf c$ such that their
time derivatives $\dot{\bf c}$ vanish in the 
(collection of) two-body Sun-object system(s). 
For any function ${\bf c}={\bf c}({\bf r},\dot{\bf r},t)$,
where $\bf r$ is the position vector and $t$ is
the time, we have
\be
\dot{\bf c}={\p{\bf c}\over\p\bf r}\dot{\bf r}+ 
{\p{\bf c}\over\p\dot{\bf r}}\ddot{\bf r}+{\p{\bf c}\over\p t}.
\ee
But since this must identically vanish in the Keplerian case, 
we are left with
\be
\dot{\bf c}={\p{\bf c}\over\p\dot{\bf r}}\tilde{\ddot{\bf r}},\label{cder}
\ee
where $\tilde{\ddot{\bf r}}$ denotes the part of acceleration
due to perturbative forces. In heliocentric coordinates,
\be
\tilde{\ddot{\bf r}}_i=\sum_{k=1 (i\ne k)}^N Gm_k\left(
{{\bf r}_{ik}\over r_{ik}^3}-{{\bf r}_k\over r_k^3}\right),
\ee
where $\tilde{\ddot{\bf r}}_i$ is the perturbative
acceleration of object $i$, $N$ the number of objects,
$G$ the gravitational constant,
$m_k$ are the objects' masses, ${\bf r}_{ik}\equiv
{\bf r}_k-{\bf r}_i$, and ${\bf r}_k$ are the heliocentric position
vectors. 

Obviously we want to choose the elements $\bf c$ such that the
partial derivatives $\p\bf c/\p\dot{\bf r}$ can be easily computed;
also, the transformation between $\bf c$ and $(\bf r,\dot{\bf r})$
should be as simple as possible so that $\dot{\bf c}$ can be viewed
in the form
\be
\dot{\bf c}=F\lbrack{\bf r}({\bf c},t),{\dot{\bf r}}({\bf c},t)
\rbrack=F({\bf c},t),\label{diffeq}
\ee
i.e., in the standard form suitable for a multitude of integration schemes
(note that we formulate the problem for single-integration methods
instead of double-integration ones).
The usual Keplerian orbital elements will not do as they are
extremely cumbersome and, worst of all, lead to formulas that have
singularities 
when eccentricity or inclination goes to zero. A practical choice is 
simply to derive from $(\bf r,\dot{\bf r})$ the Cartesian velocity and 
position the object would have at a given point in its Keplerian orbit if
the perturbing forces were removed, and use them
as the elements. This practicality is caused by the fact that 
$(\bf r,\dot{\bf r})$ at any point in a Keplerian orbit can easily be derived
from those at any other point
with the aid of the so-called $f$ and $g$
functions by Gauss (see Sect.\ 4 and eq.(\ref{fgorig}); 
also Danby \cite{Da87}). 
This makes the necessary transformations simple, and all
quantities are well defined at all times.

Depending on the situation and the choice of the integration method, we
use either
\be
{\bf c}=({\bf r}_0,\dot{\bf r}_0)_K
\ee
at a given time $t_0$ (the subscript $K$ emphasizes that the values
are to be evaluated along the imagined Keplerian orbit) or
\be
{\bf c}=({\bf r}_0,\dot{\bf r}_0,t_0;y_0=0, x_0>0)_K,
\ee
i.e., the position and velocity as well as the time at the point where
the object would last have crossed (or would next
cross) the $xz$-half plane if it moved in an
unperturbed Kepler-orbit. In both cases
we have six elements that can be used just like the traditional
osculating geometric ones. Provided that the eccentricity is never
small, one can also use the more traditional
\be
{\bf c}=({\bf r}_0,\dot{\bf r}_0,t_0;\dot r_0=0,\ddot r_0>0)_K,\label{peri}
\ee
where $({\bf r}_0,\dot{\bf r}_0)$ can be given in
spherical coordinates, and $t_0$
is now the time of perihelion. This is useful for highly eccentric
osculating ellipses that can occasionally open into hyperbolae.

\subsection{Variable time step}

A basic method of controlling the length of the time step is to
use the standard procedure of
extending phase space by introducing a new independent variable
$\tau$, related to the time $t$ by the differential equation
\be
{{\rm d}t\over{\rm d}\tau}=g({\bf c},t),\label{timeq}
\ee
where $g$ is any given function (e.g., proportional to velocity)
that can depend explicitly on time as well. The differential
equations (\ref{diffeq}) are now replaced by
\be
{{\rm d}{\bf c}\over{\rm d}\tau}=g({\bf c},t)\,F({\bf c},t),
\ee
while the time corresponding to $\tau$ is obtained by integrating
(\ref{timeq}). The stepsize for $\tau$ can be constant. 

\subsection{Variational equations}

To obtain either the first Liapunov exponent or the state transition
matrix (called matrizant in Danby \cite{Da87}; also see, e.g.,
Mikkola \& Innanen \cite{MiIn99}), one needs to 
compute the evolution of the differences
${\bf d}_{\bf c}$ between the values of ${\bf c}$ for two initially
close orbits. This is governed by the variational equations
\be
\dot{\bf d}_{\bf c}=M{\bf d}_{\bf c},\label{liap}
\ee
where the matrix $M=\p\dot{\bf c}/\p{\bf c}$ 
(see, e.g., Lichtenberg \& Lieberman \cite{Li91}). In our case
it assumes the form
\be
M_{ij}={\p^2 c_i\over\p{\bf w}\p\dot{\bf r}}\tilde{\ddot{\bf r}}
{\p{\bf w}\over\p c_j}+{\p c_i\over\p\dot{\bf r}}
{\p\tilde{\ddot{\bf r}}\over\p{\bf r}}{\p{\bf r}\over\p c_j},\label{matrix}
\ee
where ${\bf w}\equiv({\bf r},\dot{\bf r})$ (the compact vector notation
in (\ref{matrix}) is somewhat unorthodox but rather obvious). The
differences ${\bf d}_{\bf c}$ can be computed along with ${\bf c}$
using the same integration method. By definition, the differential 
equation (\ref{liap}) describes infinitesimal quantities: 
its linearity makes ${\bf d}_{\bf c}$ scale-free. The first
Liapunov exponent $\sigma$ is formally defined as
\be
\sigma=\lim_{t\rightarrow\infty} {1\over t}\ln {\vert {\bf d}_{\bf c}\vert
\over\vert {\bf d}_{\bf c}(0)\vert}.
\ee
Rather than integrating the full
variational equations, the evolution of the variations
can also be computed by
differentiating the orbit integration algorithm directly (Mikkola \&
Innanen \cite{MiIn99}).

It is perhaps worth noting that the Liapunov exponent is usually defined
for autonomous systems. If there is explicit time-dependence, one should
in principle perform the above trick of extending phase space. If 
the expansion is trivial, i.e., $g\equiv 1$ in (\ref{timeq}) to make
the system formally autonomous, we have
$\dot d_t=0$, so
we can set $d_t=0$ and ignore it.

\section{Fixed $t_0$ and low-order methods}

The partial derivatives $\p\bf c/\p\dot{\bf r}$ 
in (\ref{cder}) become trivial (either 1 or 0) if
we choose to employ $t_0=t$, i.e., a point in the actual orbit is 
also taken to be its own reference point in the Kepler orbit. Thus
\be
{{\rm d}{\bf r}_0\over {\rm d}t}=0,\quad
{{\rm d}\dot{\bf r}_0\over {\rm d}t}=\tilde{\ddot{\bf r}}.
\ee
(Note carefully that ${\rm d}{\bf r}_0/{\rm d}t\ne\dot{\bf r}_0$ -- the two
quantities are fundamentally different.)
This minimizes the work at each point and shifts the computational
load to the Keplerian transformations (via the $f,g$-functions)
needed to bring other points to the
same $t_0$-frame. Note that
${\bf r}_0$ only changes by such transformations and not by
a differential equation. For example, a naive first-order Euler step
consists of a `drift' in the Kepler
part to bring the previous point to the new frame
followed by a `kick' from the perturbative part.

The matrix $M$ in (\ref{matrix}) 
becomes especially simple in this approach as the
double derivatives vanish and the only remaining nontrivial
derivatives are $\p\tilde{\ddot{\bf r}}/\p{\bf r}$. Just like ${\bf r}_0$,
${\bf d}_{{\bf r}_0}$ evolves only by transformation from
one frame to another (i.e., not by
integration). Since ${\bf d}_{\bf c}$ represent infinitesimal
quantities (differentials), the transformation is given by
\be
\bar{\bf d}_{{\bf r}_0}={\p\bar{\bf r}_0\over\p{\bf c}}{\bf d}_{\bf c},
\ee
where the bar over a quantity denotes its value in the frame
associated with the new point; the derivatives are readily
obtained from the $f,g$ functions
(see Sect.\ 4). This transformation thus retains
the scale-free linearity of ${\bf d}_{\bf c}$ in the differential
equation.

For higher orders one usually employs multistep schemes
to keep the force evaluations at minimum; in any
case, about $n$ points at different epochs are typically needed
to construct an $n$th-order integrator. 
One would thus require about $n$ transformations per step 
in the $t_0=t$-approach, so the computational overhead from the
$f,g$-calculations quickly neutralizes the gain from the simple
derivatives as the order of the method increases.
Therefore the $t_0=t$-frame is suitable only for low-order methods such
as the modified midpoint method in a Bulirsch-Stoer integrator
(Press et al.\ \cite{Pr89}). Since
the errors in integration -- especially the energy error in conservative 
systems -- are best kept small by using a high-order
integrator, long integrations are not very suitable for the
$t_0=t$-frame. Thus the simple formulation for the variational equations
is mainly applicable to the state transition matrix (for obtaining
nearby trajectories) during relatively short integration times.

\section{Fixed $y_0$ and higher-order methods}

A common reference frame for all points gets rid 
of the excessive $f,g$-transformations.
Setting $t_0=0$ for all points is not a practical way to establish
such a frame unless the total integration time is very short.
This is because errors will be measured against just one
orbital cycle rather than the whole time span as Kepler's equation 
is solved: what is small relative to, say, $2000\pi$ is
not insignificant relative to $2\pi$. Since we would have to perform
coordinate transformations over many orbital cycles,
the slightest errors would be exponentially amplified and finally blow
up. Using a new $t_0$ every now and then to prevent transformations
over long time spans would not be very practical either, for then
the points would be only `piecewise' in a common frame of reference,
and one would have to do extra transformations near the interfaces
of the pieces to establish continuity. 

If we choose ${\bf c}=({\bf r}_0,\dot{\bf r}_0,t_0;y_0=0, x_0>0)_K$
as the common frame, we will have no difficulties
as we will not have to do transformations over more than one cycle if
we do not want to. Now $t_0$'s role 
among the new elements resembles that of the time of perihelion among
the traditional ones. Since no `perihelion' is needed now, this frame
is especially suitable for orbits at low eccentricities.

An expression for $t_0$ can be found using
the $f$ and $g$ functions. Using the subscript 0 for these functions
to emphasize that we shift from $({\bf r},\dot{\bf r})$ to
$({\bf r}_0,\dot{\bf r}_0)$, we have
\be
{\bf r}_0= f_0{\bf r}+g_0{\dot{\bf r}},\label{rtransf}
\ee
where 
\bea
f_0&=&1-{a\over r}(1-\cos\hat E),\\
g_0&=&\Delta t-{a^{3/2}\over\sqrt{\mu}}(\hat E-\sin\hat E),\label{gdef}
\eea
and $\mu=G(m_i+m_0)$, $a=-1/\alpha$, $\alpha=\vert\dot{\bf r}\vert^2
/\mu-2/r$. Also, $\Delta t=t_0-t$ and $\hat E$ is 
the corresponding difference
between the eccentric anomalies at ${\bf r}_0$ and ${\bf r}$ (note
that the absolute values of the eccentric anomalies are never needed
and that $\hat E$ is always well defined while $E$ is not).
Solving for $\Delta t$ in the difference-formed Kepler's equation 
(see, e.g., Danby \cite{Da87})
\be
{\sqrt{\mu}\over a^{3/2}}\Delta t=\hat E+{u\over\sqrt{\mu a}}
(1-\cos\hat E)-s\sin\hat E,\label{kepler}
\ee
where $u={\bf r}\cdot\dot{\bf r}$ and $s=1+\alpha r$,
we obtain
\be
g_0={u a\over\mu}(1-\cos\hat E)+r\sqrt{{a\over\mu}}\sin\hat E.
\label{newg}
\ee
Since we have set $y_0=0$, we know that
\be
f_0 y+g_0\dot y=0,
\ee
so, using the new form (\ref{newg}) for $g_0$, we have
\be
y+A(1-\cos\hat E)+B\sin\hat E=0,
\ee
where
\be
A={u a\over\mu}\dot y-{a\over r} y,\quad B=r\sqrt{{a\over\mu}}\dot y.
\ee
Solving for $\sin\hat E$ and $\cos\hat E$
(and requiring that $x_0>0$ when $y_0=0$), we finally obtain 
$\hat E$ from
\be
\sin\hat E={-B(A+y)+A\sqrt{B^2-2Ay-y^2}\over A^2+B^2}
\ee
and
\be
\cos\hat E={A(A+y)+B\sqrt{B^2-2Ay-y^2}\over A^2+B^2}.
\ee
The above formulae hold for prograde motion $(L_z>0)$; if the
motion is retrograde, the branch sign immediately in front of the square
root terms is changed to negative
(note that $\hat E$ always has the same sign as $\Delta t$). 
$A^2+B^2$ as well as the
square root are positive definite (for elliptic motion), so the
formulae hold everywhere. Substituting $\hat E$ to (\ref{kepler}) 
or to $g_0$ in (\ref{gdef}), we get $\Delta t$ and thus
\be
t_0=t+\Delta t.
\ee
From (\ref{rtransf}) we obtain ${\bf r}_0$, 
while $\dot{\bf r}_0$ is found from
\be
\dot{\bf r}_0= \dot f_0{\bf r}+\dot g_0 \dot{\bf r},\label{rdottransf}
\ee
where
\bea
\dot f_0&=&-{\sqrt{\mu a}\sin\hat E\over r r_0},\\
\dot g_0&=&1-{a\over r_0}(1-\cos\hat E).
\eea

The integration procedure is now as follows:
From the initial values of $({\bf r},\dot{\bf r})$ we get the
corresponding ${\bf c}=(t_0,{\bf r}_0,\dot{\bf r}_0)$ to
be integrated with whatever numerical method we have chosen.
The derivatives $\dot{\bf c}$ are given by (\ref{cder}); 
using (\ref{rtransf}) and (\ref{rdottransf})
we can write $\dot{\bf c}$ as
\be
{{\rm d}\hat{\bf r}_0\over {\rm d}t}=
\left({\p\hat f_0\over\p\dot{\bf r}}\cdot\tilde{\ddot{\bf r}}
\right){\bf r}+\left(
{\p\hat g_0\over\p\dot{\bf r}}\cdot\tilde{\ddot{\bf r}}\right)
\dot{\bf r}+\hat g_0 \tilde{\ddot{\bf r}}
\ee
and
\be
{{\rm d}t_0\over {\rm d}t}={\p t_0\over\p\dot{\bf r}}\cdot\tilde{\ddot{\bf r}},
\ee
where the hat over $f_0,g_0,{\bf r}_0$
is either uniformly read as a dot or ignored everywhere.

The values of ${\bf r},\dot{\bf r}$ needed in computing
the derivatives (and thus obtained as `by-products') 
are given by
\be
{\bf r}=f {\bf r}_0+g \dot{\bf r}_0,\quad
{\dot{\bf r}}=\dot f {\bf r}_0+\dot g\dot{\bf r}_0,\label{fgorig}
\ee
where $f,g$ are defined as $f_0,g_0$ above but, of course, with
$({\bf r},\dot{\bf r})$ and $({\bf r}_0,\dot{\bf r}_0)$ interchanged
everywhere. $\hat E$ is in this case obtained by solving Kepler's 
equation (\ref{kepler}) (with $\Delta t=t-t_0$). One should used the fast,
quartically convergent iteration technique (Danby \cite{Da87}) that will
require only a couple of iterations. 
A suitable initial guess is, e.g.,
$\hat E_0=\hat E_{\rm prev}+\sqrt{\mu} a^{-3/2}(t-t_{\rm prev})$, 
`prev' referring to the previous point.

One must not let the integrated $t_0$ fall too much behind the
actual time $t$ lest the errors accumulate; this is prevented
by adding
multiples of periods to an old $t_0$ in the same way as one
would set a new perihelion time every now and then even if all the
other elements remained unchanged
(or by finding a `fresh' $t_0$ regularly during the integration from
some obtained $({\bf r},\dot{\bf r})$). This does not disturb the
calculations in any way, and the elements of all points are always
in the same frame of reference.

If the osculating perihelion is always well defined, we can use
the frame (\ref{peri}) and replace $\hat E$ by the absolute
value of eccentric anomaly so that $t_0$ becomes the time of perihelion. 
When working with highly eccentric
orbits, the `universal formulation' via Stumpff functions (Danby \cite{Da87})
may be easier than using trigonometric/hyperbolic functions.

\section{Numerical examples}

We discuss here simple numerical examples that illustrate the basic
properties of the perturbative formulation. 
It is not in the scope of this short study
to present a detailed analysis of some chosen system: the main emphasis is on
the choice of coordinates that represent a near-integrable system, not on
the technical details of the integration method used.
 
As an example of the $y_0=0$-frame, a multistep method, 
and a dissipative force, we consider a satellite
experiencing a drag force $-k\dot{\bf r}$ (for orbits of low
eccentricity, this frame is not just useful but necessary
when multistep methods are used). With the perturbative scheme,
the spiralling orbit can obviously be computed with considerably 
fewer steps than in the non-perturbative approach. An interesting quantity
is the magnitude of this advantage as a function of the coefficient $k$.
We used a basic Adams-Bashforth multistep method to
integrate the orbit with different values of $k$ and required tolerance.
Regardless of the tolerance and the order of the method,
the ratio of the stepsize in the perturbative approach to that
in the direct one was typically an exponential function of $\log_{10} k$.
For example, with a fifth-order method and an accuracy of one part in 
$10^{11}$, the perturbative scheme could use a 50-100 times longer step
at $k=10^{-6}$ ($k$ scaled to be 
roughly descriptive of the proportional strength of the 
perturbation), while at
$k=10^{-4}$ the ratio was about 20, and even at $k=0.01$ the
steps could be some five times longer. 
A rough rule of thumb for
the stepsize ratio $\cal R$ (at $k\le 0.01$) is
\be
{\cal R}\approx 10\times  2^{-(3+\log_{10} k)}.
\ee
Thus the benefits of the near-integrable approach
are clear even when the perturbation is no longer small and the equations
of motion are far from analytically integrable. Using $g$ in (\ref{timeq})
does not really alter the ratio of the steplengths since much the same $g$
can be used for both the perturbative and the direct approach 
(in the above case, 
a practical value for $g$ is inversely proportional to the object's speed
or even its angular speed).

As an example of the $t_0=t$-frame, we computed orbit values
for asteroids in eccentric orbits using the Bulirsch-Stoer extrapolation
scheme with the modified midpoint method (Press et al.\ \cite{Pr89}). 
As in the previous example, we used
varying perturbation strengths (in some cases `Jupiter' was tens of times 
more massive than in reality) and tolerance levels. In this case, the
near-integrable/direct stepsize ratio
had no clear correlation with these (except for
very small perturbations and low accuracies, of course); this was mostly
due to the strongly varying strengths of perturbations as the asteroid
proceeded in its orbit.
We found that, on average, the timestep required for the near-integrable
approach can be some five times longer than in the direct method. 
Similar results are obtained with spatially fixed frames and multistep
methods. For
orbits not close to giant planets (i.e., typically
at relatively low eccentricities or high inclinations),
the stepsize ratio can be about ten; the rule of thumb above seems to
apply rather generally. Even though the steps take a longer
time to compute in our approach due to the Keplerian transformations, 
the overall speed of the algorithm is usually noticeably faster than
that of direct integration.

Eccentric orbits bring about the problem of close encounters.
One can, in principle, switch to a new dominating central body whenever
necessary and choose a suitable near-integrable Keplerian configuration. 
However, we have found that all of the time during a `moderately
close' encounter is spent in the transition zone where no
single body dominates the gravitational field strongly enough for the
near-integrable approach to be efficient. For an asteroid
approaching Jupiter, this zone
is roughly between 0.02 AU and 1 AU from the planet.
`Proper' close encounters are thus very
rare, and even in them the time spent in the transition zone dominates the
computational effort. A further problem is that the object may
wander in and out of the transition zone several times during the
encounter. Close encounters are thus clearly easiest to handle in
conventional integration variables. Whenever the integrator finds that
the strength of the perturbation gets too large -- typically well over
a hundredth part of the dominating force -- it switches to `ordinary mode'.
Frequent close encounters undermine the efficiency of
near-integrability, so only an object not engaged 
in continuous interplay with Jupiter can be integrated well in this scheme.

To summarize,
our numerical tests show that the near-integrable approach is efficient
whenever the perturbations do not considerably exceed about a hundredth
part of the force caused by the dominating body. Occasional close encounters
are no obstacle, but they must be handled with a separate method. Thus
the impact of this approach should be greatest in integrating 
orbits (perturbed also by dissipative forces) that mostly stay away from the
transition tori surrounding the orbits of the giant planets. Orbits with
high eccentricities can be efficiently integrated if their inclinations
are suitable.

\section{Conclusions}

The primary motivation for our study was to find out whether there is a
way of formulating a near-integrable non-symplectic scheme  
more efficient than the direct N-body computation. The traditional
perturbative methods are now {\it pass\'e}, but some of their basic 
principles can still be used for efficient numerical computation.
Since the straightforward
formulation presented here is based on the choice of
the integrated variables, the actual integration method can be chosen
quite freely. The simplest frame is the one with $t_0=t$, best used
with low-order extrapolation methods such as Bulirsch-Stoer;
the longest integration steps are allowed by high-order
multistep schemes in a spatially fixed frame.

The error in integration depends on the specific method chosen; not much
can be said about the error elements introduced by the generic
principle. Various numerical experiments indicate that
(e.g., in the case of our solar system)
the stepsize can be sizably larger in this approach than
in a non-perturbative one to cause similar error magnitudes
in the two methods. The main limitation is, of course, the strength of the
perturbation: roughly speaking, its maximum value for efficient use
is of the order of 
one percent of the dominating force.

Integrators based on near-integrability are somewhat more
efficient than direct ones, but this advantage is not as clear as that
provided by symplectic integrators especially in long-term integrations.
The special characteristics exhibited by SIs in symplectic systems are,
indeed, quite remarkable, and due to a `deeper' connection
with the dynamics of the system than mere near-integrability and
conservation of energy.
However, when SIs cannot be used, the next best thing to do to maintain
some knowledge about the nature of the system may often be to use
the near-integrable formulation.
In some cases (especially in dissipative systems) 
the advantage gained can be considerable.

\end{document}